# Fast Optical Layer Mesh Protection Using Pre-Cross-Connected Trails

*Timothy Y. Chow[1], Fabian Chudak, and Anthony M. Ffrench*

**Abstract.** Conventional optical networks are based on SONET rings, but since rings are known to use bandwidth inefficiently, there has been much research into *shared mesh protection*, which promises significant bandwidth savings. Unfortunately, most shared mesh protection schemes cannot guarantee that failed traffic will be restored within the 50 ms timeframe that SONET standards specify. A notable exception is the p‑cycle scheme of Grover and Stamatelakis. We argue, however, that p‑cycles have certain limitations, e.g., there is no easy way to adapt p‑cycles to a path-based protection scheme, and p‑cycles seem more suited to static traffic than to dynamic traffic. In this paper we show that the key to fast restoration times is not a ring-like topology *per se*, but rather the ability to *pre-cross-connect* protection paths. This leads to the concept of a *pre-cross-connected trail* or *PXT*, which is a structure that is more flexible than rings and that adapts readily to both path-based and link-based schemes and to both static and dynamic traffic. The PXT protection scheme achieves fast restoration speeds, and our simulations, which have been carefully chosen using ideas from experimental design theory, show that the bandwidth efficiency of the PXT protection scheme is comparable to that of conventional shared mesh protection schemes.

## 1. INTRODUCTION

Modern optical networks carry large amounts of traffic, so it is imperative that they be able to survive an accidental failure such as a fiber cut or a node failure. Traditionally, such survivability has been provided by means of *ring protection*, e.g., SONET unidirectional (UPSR) or bidirectional (BLSR) rings. Rings can recover automatically from any single link or node failure by rerouting traffic around the other side of the ring, and their switch completion times are very low (e.g., 50 ms for a SONET BLSR with a 1200 km circumference, provided that there is no extra traffic), thereby minimizing the amount of data lost. However, rings are known to have low bandwidth efficiency. There has therefore been much research into *shared mesh protection*, which promises significant bandwidth savings (see [1]–[7] for a sample—by no means comprehensive—of the literature). These savings come partly from the flexibility of an arbitrary mesh topology, which allows many traffic demands to take more direct routes to their destinations than if they were constrained to a ring topology. More importantly, however, bandwidth savings are achieved through *sharing*. The basic assumption behind sharing is that failures in optical networks are rare enough that the probability of two independent failures occurring simultaneously is negligible. Therefore, if two light paths that carry working traffic do not have any common point of failure, then their protection paths can share the same unit of bandwidth, since they will never request that unit of backup bandwidth simultaneously. Because of sharing, the amount of spare capacity in a mesh-protected network can be as little as 60% (or even less) of the working capacity, whereas in a BLSR or UPSR this fraction is always at least 100%.

Most shared mesh protection schemes can be classified as either *link-based* or *path-based*. We define these terms and compare them in the next section; for now it suffices to observe that link-based and path-based schemes each have their respective pros and cons. Depending on a particular network's characteristics and needs, one or the other approach may be preferable.

---
[1] 250 Whitwell Street #2, Quincy, MA 02169, USA

Shared mesh protection looks good on paper, but as practical implementations have begun to be built and tested, certain difficulties have surfaced. One of these has been the experimental fact that shared mesh protection, whether link-based or path-based, typically does not restore traffic as quickly as SONET rings do, and therefore runs the risk of being unable to guarantee the quality of service that is typically required for high-priority traffic. The only obvious way to achieve very fast switch completion times in a mesh network is to give up the idea of sharing and to use *dedicated* (also known as 1+1) protection instead, but this means giving up the bandwidth savings that motivated mesh protection in the first place.

One of the few mesh protection schemes to address this quandary satisfactorily is the *p-cycle* concept of Grover and Stamatelakis [3]. We say more about p-cycles in a later section; for now let us just remark that the idea is to route the working traffic using an arbitrary mesh routing algorithm, but to constrain the protection routes to lie on certain predetermined "p-cycles" or rings. Grover and Stamatelakis report that p-cycles achieve the "speed of rings with the efficiency of mesh."

Impressive as Grover and Stamatelakis's results are, the p-cycle concept does have certain limitations. It is inherently a link-based scheme, and is therefore saddled with all the usual pros and cons of link-based schemes. More subtly, Grover and Stamatelakis have found that in many situations, achieving high bandwidth efficiency requires the deployment of *large* p-cycles. In a real network that carries live traffic, demands are not static but are *dynamic:* they come and go incrementally over time. A network provider faced with a new traffic demand that cannot be handled with existing p-cycles must therefore choose between allocating a small p-cycle that meets the current demand cheaply but that may be inefficient over the long run, or allocating a large p-cycle that requires a large investment up front and that may be wasted if expected future demands never materialize. This dilemma can be solved by starting with small p-cycles and periodically reoptimizing their size as traffic increases, but continual reoptimization entails significant management overhead.

In this paper we take a step beyond p-cycles. We argue that rings and p-cycles restore traffic quickly not because of their circular topology but because their protection routes are *pre-cross-connectable*. We are thus naturally led to consider arranging protection capacity not only into cycles but more generally into *pre-cross-connected trails* (or PXTs for short; the precise definition will be given later). Like p-cycles, PXTs achieve the "speed of rings with the efficiency of mesh," but they are more flexible than p-cycles. In particular, PXTs can be used in either a link-based or a path-based scheme, and they are well suited to dynamic traffic because they grow and shrink incrementally in a natural way.

## 2. LINK-BASED VERSUS PATH-BASED PROTECTION

As we mentioned in the introduction, some protection schemes are *link-based* while others are *path-based*. In this section we define these two terms and give a brief comparison of their respective merits.

In link-based protection, the nodes—let us call them *A* and *B*—at either end of a failed link are responsible for detecting the failure and switching the traffic from the failed link onto a protection path *P* that bypasses the failure. The failed link may be utilized by a large number of different light paths, each with a different source and destination. After the failure, these light paths travel from their source node to node *A* as before, then take the protection path *P* to get to node *B*, then finally travel from node *B* to their final destinations. SONET BLSR's use link-based protection.

In path-based protection, it is the source and destination nodes of each individual working light path that are responsible for switching the traffic onto a protection path. As in link-based protection, a single failed link may cause many different light paths to fail. However, in path-based protection, each one of these light paths is free to travel on a completely different protection path from source to destination. In particular, there is no need for any of them to visit the nodes *A* and *B* at the ends of the failed link. SONET UPSR's use path-based protection. Path-based protection is also known as *end-to-end protection*.

There is one more point that should be made about path-based protection. In principle, the source and destination of a given working light path could choose different protection paths depending on which link or node along the working path fails. Although such failure-dependent routing can improve the bandwidth efficiency slightly, it requires additional signaling to isolate the fault. This slows down the recovery process and so is rarely used in practice. In this paper we assume that the same protection path is always used for a given working path. Naturally, this means that the protection path must be disjoint from the working path.

How does one choose between link-based and path-based protection? There are several factors to consider.

   a. **Shared path-based protection tends to use less total bandwidth than shared link-based protection.** One reason is that in link-based protection, there is a backhaul problem. A protection light path may travel to node *A* and then double back on itself in order to get to node *B*.

   b. **Shared link-based protection tends to be faster than shared path-based protection.** The reason is that in link-based protection, the failure detection and repair happens locally, whereas in path-based protection the signals must travel all the way to the source and the destination. Furthermore, as already mentioned, a single fiber cut usually triggers a large number of alarms in a path-based scheme, and processing all these alarms simultaneously can bog down the network.

   c. **It is difficult if not impossible for a link-based scheme to protect against node failures.** Link-based schemes rely on the nodes on either end of a link to perform a protection switch; if one of these nodes fails, then it cannot perform the switch. A path-based scheme can simply choose node-disjoint paths from end to end for all its light paths and then node failures are automatically survivable—unless it is the source or destination node that fails, but in that case it is impossible to recover from the failure anyway. (Note, however, that we said "difficult" rather than "impossible"; for example, a SONET BLSR, which we described as a link-based scheme, *is* able to survive node failures. But this is only because a BLSR has, in addition to its basic mechanism for surviving a single link failure, a complicated system involving ring maps and squelch tables for detecting and coping with multiple failures.)

As we said before, in some situations link-based protection is the right choice and in other situations path-based protection is preferable. Ideally, a comprehensive protection methodology should be flexible enough to provide either link-based or path-based protection, depending on the needs of an individual network.

## 3. PRE-CROSS-CONNECTION: THE KEY TO FAST PROTECTION

### 3.1. Branch points

Why does a conventional shared mesh protection scheme experience a slower restoration time than a SONET BLSR? The key observation is that that in a BLSR, only the nodes on either side of a failure need to make a real-time switch. The rest of the protection path is *pre-cross-connected*, so that the intermediate nodes on the protection path simply pass through the traffic without having to make a switching decision. In contrast, in a shared mesh environment, every intermediate node along the protection path may have to make a real-time switch. This adds considerable delay to the protection switching time.

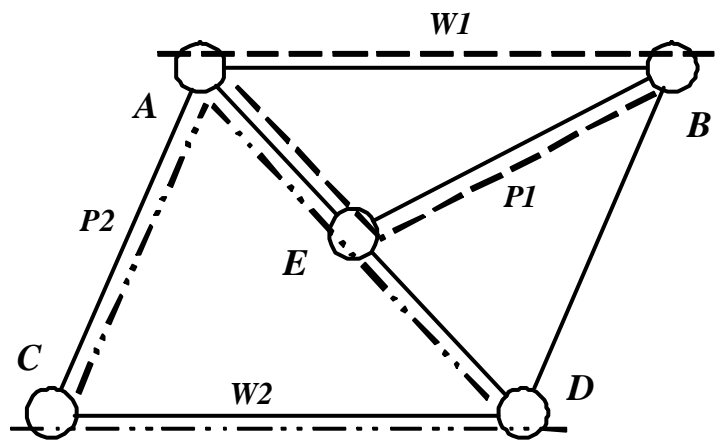

**Figure 1**

Figure 1 may clarify this point. Solid lines represent physical links while broken lines represent light paths. The working path *W1* (from *A* to *B*) and the working path *W2* (from *C* to *D*) have no common point of failure, so let us assume that their protection paths *P1* (from *A* to *E* to *B*) and *P2* (from *C* to *A* to *E* to *D*) share protection bandwidth on the link *AE*. Now, if *W1* fails, then node *E* must connect *AE* to *EB*, whereas if *W2* fails, then node *E* must connect *AE* to *ED*. Therefore, in order for node *E* to decide which connection to make, it must be informed of the location of the failure, and this information is available only *after the failure has occurred* and cannot be pre-calculated. Although one could arbitrarily decide to pre-cross-connect *AE* to either *EB* or *ED*, neither choice would entirely eliminate the need for a real-time switch at *E*; if *AE* is pre-cross-connected to *EB*, then *E* must make a real-time switch when *W2* fails, and if *AE* is pre-cross-connected to *ED*, then *E* must make a real-time switch when *W1* fails.

It is convenient to introduce some terminology to describe the above situation. We define a *branch point* to be a node *X* with the property that, no matter how the protection capacity is pre-cross-connected at *X*, there exists a failure scenario for which some needed protection path that has *X* as an intermediate node is not properly pre-cross-connected at *X*. Notice that branch points can arise regardless of whether one uses link-based or path-based shared mesh protection. Each branch point along a protection path can add several milliseconds to the total switch completion time.

These observations explain why Grover and Stamatelakis's mesh protection scheme [3] succeeds in achieving ring-like speeds. As we mentioned before, in their scheme, all the protection capacity is organized into certain predetermined rings or p-cycles, and therefore no troublesome branch points can arise. The p-cycles are pre-cross-connected just as in a BLSR. When a failure occurs, the nodes at either end of the failure must react and perform a real-time switch, but all the intermediate nodes on the protection path simply pass through the traffic.

We now come to the main insight of the present paper. *Although a ring or a p-cycle has no branch points, the converse is not true.* That is, if there are no branch points, this does not automatically imply that the protection bandwidth is arranged into a set of rings. As an example, suppose we modify Figure 1 slightly by choosing a different route for *P1*, as shown in Figure 2.

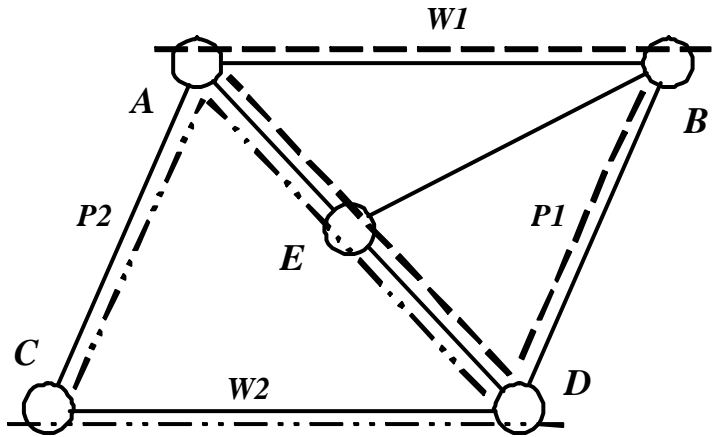

**Figure 2**

On both *AE* and *ED*, *P1* and *P2* share bandwidth. We pre-cross-connect *CA* to *AE*, *AE* to *ED*, and *ED* to *DB*.

The crucial observation is that the branch point at node *E* has now been removed, so that pre-cross-connecting *AE* to *ED* works correctly for all failure scenarios. Furthermore, despite appearances, *D* is *not* a branch point either. To see this, note first that if *W1* fails, then the pre-cross-connection between *ED* and *DB* is exactly what we want, because it connects up *P1* and allows the intermediate node *D* to pass through protection traffic without making a real-time switch. On the other hand, if *W2* fails, then the *ED-DB* pre-cross-connection is admittedly improper and must be broken in order to terminate *P2* at *D*, but the key fact is that for *this* failure, node *D* is not an *intermediate* node on the protection path but an *endnode*. (Recall the proviso about intermediate nodes in the definition of a branch point.) The idea is that *D* must perform a real-time switch in any case, and no *additional* delay is incurred by requiring it to break the cross-connection between *ED* and *DB*. Similarly, *A* is not a branch point. In Figure 2, just as in a ring, regardless of the failure, only the endnodes need to switch.

We argue, therefore, that as long as we avoid branch points, we can achieve the "speed of rings." Moreover, since avoidance of branch points can be achieved without necessarily committing to rings or p-cycles, we open up the possibility of overcoming some of the limitations of the latter. For example, we are no longer committed to a link-based protection scheme, and even if we do choose a link-based scheme, we can use our newfound flexibility to improve the bandwidth efficiency or the robustness to dynamic traffic. These points will be explored further in the following sections.

### 3.2. Graph-theoretic terminology

Before we proceed further it is helpful to give a formal statement of the problem we are trying to solve. To do this we must first briefly review some graph-theoretic terminology.

By a *graph* we mean an *undirected multigraph*, i.e., an undirected graph that may have multiple edges. An edge is also called a *link connection*. A *link* in a graph *G* is the set of all edges between a given pair of nodes. Two edges that share exactly one endnode *v* in common are said to be *incident to each other at v*. The *degree* of a node *v* is the number of edges that have *v* as an endnode. A graph is *regular* if every node has the same degree.

In this paper, the distinction between an *edge* and a *link* is more important than is usually the case, so we explain it in slightly more detail. The reader familiar with ITU terminology may find it helpful to think of the terms "link" and "link connection" as short for "regenerator section link" and "regenerator section link connection" (although this correspondence should not be taken *too* literally since we describe our algorithms in terms of abstract graphs and not in terms of actual optical networks). Alternatively, the reader can think of an edge as the smallest unit of bandwidth that can be switched, while a link comprises the totality of bandwidth between two adjacent nodes. For example, in a WDM network (with no sub-wavelength muxing), an edge would be a wavelength, while a link would comprise all the fibers in the conduit between two nodes.

In the graph in Figure 3 below, there are two edges or link connections on the link between nodes *D* and *E*, namely *e* and *f*. The degree of node *D* is 4.

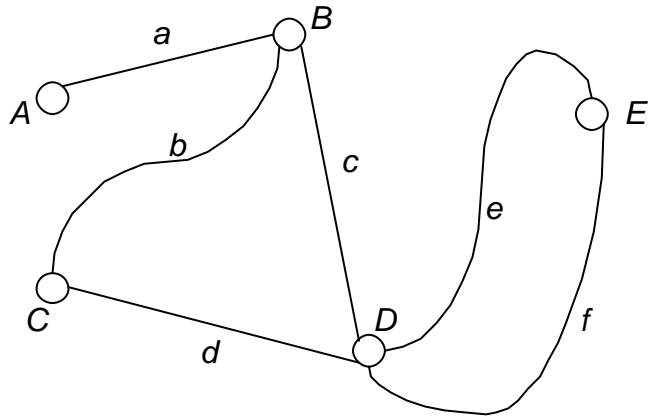

**Figure 3**

A *walk* in $G$ is an alternating sequence of nodes and edges $(v_0, e_1, v_1, e_2, v_2, \ldots, v_{n-1}, e_n, v_n)$ in $G$ such that for all $i$, the endnodes of $e_i$ are $v_{i-1}$ and $v_i$. A *trail* is a walk whose edges are all distinct and a *path* is a walk whose nodes are all distinct. A walk is *closed* if $v_0 = v_n$. Note that a path cannot be closed unless $n = 0$. A walk is said to *connect* $v_0$ and $v_n$.

In Figure 3, $(C, d, D, f, E, e, D, d, C)$ is a walk but not a trail because the edge $d$ is repeated. On the other hand, $(C, d, D, f, E, e, D, c, B, b, C)$ is a trail—in fact a closed trail—but not a path, because the node $D$ is repeated.

Two walks in $G$ are *edge-disjoint* (respectively, *link-disjoint*) if there is no edge (respectively, link) that both of them traverse. The *interior* of a walk is the set $\{v_1, v_2, \ldots, v_{n-1}\}$, i.e., the set of all of its nodes other than its endnodes. Two walks are *node-disjoint* if they are link-disjoint and no node in the interior of one walk is a node in the other walk.

For example, in Figure 3, the paths $(D, e, E)$ and $(D, f, E)$ are edge-disjoint but not link-disjoint, and hence they are not node-disjoint. The paths $(A, a, B, c, D)$ and $(C, b, B)$ are not node-disjoint, because $B$ is an interior node of the first path that is also a node of the second path. On the other hand, the paths $(A, a, B, b, C)$ and $(E, e, D, d, C)$ *are* node-disjoint, because even though node $C$ is on both paths, $C$ is not an *interior* node of either path.

### 3.3. Problem statement

For brevity we describe only a path-based protection scheme and omit the details of the analogous link-based version. We also assume that all traffic and all links are bidirectional.

**INPUT.** A graph $G$ and a list $D$ of *demands*. A demand $d$ is an unordered pair of distinct nodes—the *terminals* of $d$—of $G$; the same demand may appear multiple times in the list $D$. For simplicity we assume that each edge of $G$ has unit cost and unit capacity and that each demand in $D$ requires unit capacity to route, although allowing arbitrary integer costs and capacities would not change the essence of the problem.

**OUTPUT.** An *allocation plan*, i.e., a list that contains, for each demand *d* in *D*, two paths [note: paths, not arbitrary trails or walks] connecting the terminals of *d*—a *working path* and a *protection path*— and that satisfies the following conditions:

   a. For each *d*, the working and protection paths are node-disjoint. (Note: here and in what follows, "node-disjoint" may be changed to "link-disjoint" if survivability against node failures is not required.)

   b. No edge [note: edge not link] that appears in the working path of some demand appears in either the working or protection path of any other demand.

   c. If the working paths of two demands $d_1$ and $d_2$ are not node-disjoint, then their protection paths are edge-disjoint.

   d. There are no branch points. That is, if *v* is a node and $e_1$, $e_2$, and $e_3$ are three distinct edges that each have *v* as an endnode, and the protection path of some demand contains both $e_1$ and $e_2$, then no protection path of any demand contains both $e_1$ and $e_3$.

The number of edges that appear in some working (respectively, protection) path is called the *working* (respectively, *protection*) *bandwidth* of the allocation plan. Note that some protection edges are *shared*, i.e., they appear in the protection paths of more than one demand; such edges are counted only once when computing the protection bandwidth. The *total bandwidth* is the sum of the working and protection bandwidths. The smaller the total bandwidth, the better the allocation plan. The objective is to find as good an allocation plan as possible.

This completes the problem statement. Condition d, on branch points, is what distinguishes our shared mesh protection problem statement from others in the literature.

We are now able to explain the "PXT" terminology alluded to earlier. Given an allocation plan, we say that two edges $e_1$ and $e_2$ are *pre-cross-connected* if they appear consecutively in some protection path. Pre-cross-connection should be thought of as "linking up" the protection edges. The crucial fact, easily proven, is that although condition d does not force the protection edges to be arranged into a disjoint union of cycles, *the absence of branch points does imply that the protection edges must be pre-cross-connected into a disjoint union of trails* (some of which may be closed trails). We call each such trail of protection edges a *pre-cross-connected trail* or *PXT* of the allocation plan. Note that each protection path is part of some PXT, but that a PXT may be much longer than any single protection path, since protection paths may overlap (e.g., in Figure 2 the two protection paths combine to form a PXT of length four). In fact, a PXT may even form a closed trail in which the last edge comes back and is pre-cross-connected to the first edge. Such a PXT is called a *closed PXT*. A closed PXT is similar to a ring, except that a closed PXT may "self-intersect" in the sense of visiting the same node or link more than once.

As we argued above, the absence of branch points allows us to achieve ring-like protection speeds, but two important questions remain. (1) Given a graph and a list of demands, how does one compute a good allocation plan in a reasonable amount of time? (2) Is it still possible to achieve low total bandwidth when the no-branch-point constraint is imposed? These questions are addressed in the next sections.

## 4. COMPUTING AN ALLOCATION PLAN

Now that we have a formal statement of our PXT protection scheme as a combinatorial optimization problem, our first inclination might be to apply a standard mathematical programming or local search algorithm to try to find solutions to given instances. But although we hope to pursue such approaches in the future, we do not discuss them further in the present paper. The reason is that most such algorithms are best suited to so-called *offline* computation, by which we mean that the list of demands is completely known in advance and does not change with time. While this is a reasonable assumption in some networks, most real networks experience *dynamic traffic*, i.e., the list of demands arrives over time. Each demand must be satisfied when it arrives, without knowing what the future may bring. Moreover, network operators typically require a *cap-and-grow* routing method, i.e., once a demand is routed, its routing should if possible remain undisturbed when later demands arrive and are routed. In other words, re-routing and re-optimizing existing demands when a new demand arrives is strongly discouraged if not forbidden. Under these circumstances, what is needed is an *online algorithm* that routes one demand at a time. In spite of their practical importance, online algorithms and their *competitive ratios*—i.e., their performance relative to offline algorithms—have not received much attention in the literature on shared mesh protection. We hope that our work here will be a first step towards remedying this deficiency.

Suppose that we have a graph $G$ and a list $D$ of demands for which an allocation plan has already been calculated. Now suppose a new demand $d = \{u, v\}$ arrives and we need to find a working path $w(d)$ and a protection path $p(d)$ for it, without disturbing the existing allocation plan. The idea is to find the working path first, and then to find the protection path that re-uses existing protection bandwidth as much as possible, subject to the constraint that the protection path be a path, i.e., that it never revisit the same node more than once. (This requirement that the protection path be a path complicates the algorithm more than one might initially expect, as the reader will see shortly. We say a few words at the end of this section about relaxing this requirement.) More precisely, we proceed as follows.

 a. Find $w(d)$ by applying a shortest-path algorithm such as Dijkstra's algorithm to the unused part of $G$ (i.e., the edges that are not already used in the allocation plan).
 b. The existing protection edges are, as explained above, arranged into PXT's. Form a list $L_1$ of all PXT's.
 c. Discard all closed PXT's from $L_1$ except those that contain at least one occurrence of $u$ and at least one occurrence of $v$. (The reader should convince himself that unless a PXT $T$ contains both $u$ and $v$, $p(d)$ cannot possibly use any edge from $T$ without introducing a branch point.)
 d. Pick a PXT $T$ and find all occurrences of $u$ and $v$ on it. These occurrences subdivide $T$ into *subtrails*, i.e., contiguous segments of $T$ that run from one occurrence of $u$ or $v$ to the next occurrence of $u$ or $v$.
 e. Discard any such subtrails that are not paths, and append the remaining subtrails, if any, to a list $L_2$. (If any edge from a subtrail is used in $p(d)$, then the entire subtrail must be used, and since $p(d)$ is required to be a path, it cannot contain a subtrail that is not a path.)
 f. Delete $T$ from $L_1$. If $L_1$ is empty, go to step g; otherwise go to step d.

g.  Discard every path from $L_2$ with a *prohibited edge:* A prohibited edge is an edge that either has an endnode in the interior of $w(d)$ or that is contained in the protection path of an existing demand $d' \neq d$ whose working path $w(d')$ is not node-disjoint from $w(d)$.

The paths that remain in $L_2$ are called *shortcut paths*. These shortcut paths are used to help create an auxiliary graph $H$ as follows. The nodes of $H$ are the same as the nodes of $G$. If $v_1$ and $v_2$ are nodes in $G$ and there exist one or more edges $e'$ in $H$ between $v_1$ and $v_2$ such that (1) $e'$ does not appear in any existing working or protection path or in $w(d)$ and (2) $e'$ is not a prohibited edge, then we create an edge $e$ in $H$ between $v_1$ and $v_2$. (Only one such edge is created in $H$ between $v_1$ and $v_2$ even if there are many edges $e'$ in $G$ between $v_1$ and $v_2$ with the necessary properties.) We call $e$ an *unused edge*. It has a cost of one unit. Additionally, for each shortcut path $P$, we create an edge in $H$ of zero cost between the endnodes of $P$. Such edges are called *shortcut edges*.

If we now run Dijkstra's algorithm on $H$ to find the minimum-cost path between $u$ and $v$, and then "expand" each shortcut edge into the shortcut path in $G$ that it came from, then this produces the protection route $p(d)$ that uses the least amount of new bandwidth. Note that the idea behind the shortcut paths and edges is that we want $p(d)$ to consist of existing protection bandwidth as much as possible, since using such edges does not add to the overall cost of the final allocation plan. However, since branch points are forbidden, a shortcut path must be used either in its entirety or not at all. The effect of using a shortcut path is to "jump" from one end to the other at zero cost; this is modeled by the shortcut edges.

However, there is one slight problem with using Dijkstra on $H$. It is possible that when the shortest path in $H$ is "expanded" into a walk in $G$, the result will not be a path in $G$. For example, two shortcut paths may cross each other in $G$ but their corresponding shortcut edges in $H$ may not. Therefore, in order to ensure that $p(d)$ is a path, one must mark each edge $e$ in $H$ with a list of its "rivals," i.e., edges $e'$ with the property that any path containing both $e$ and $e'$ expands into a non-path in $G$. Then one must run a constrained version of Dijkstra that ensures that rival edges never appear in the same path. For details of this constrained version of Dijkstra, see the appendix.

[ **Note:** The constrained Dijkstra algorithm has the drawback of having a worst-case exponential running time, which is not very satisfactory for a practical online algorithm. In practice, fortunately, it runs rapidly on the sample networks that we have tried. However, one should note that the only reason to run constrained Dijkstra rather than ordinary Dijkstra is to ensure that the protection paths are indeed paths. If the operator of a network finds it acceptable to have protection "paths" that may revisit the same node or link more than once, then ordinary Dijkstra may be used. Not only will this speed up the algorithm, but the added flexibility of allowing such "self-intersections" can potentially increase the overall bandwidth efficiency. However, in spite of these considerations, our experimental results below cleave to tradition and enforce the constraint that protection paths must indeed be paths. ]

We remark in passing that in contrast to p-cycles, PXT's are typically not closed, and therefore can be extended incrementally at either end when new demands arrive. They also shrink incrementally when old demands disappear. This makes them well suited to online algorithms.

## 5. EXPERIMENTAL RESULTS

The problem of evaluating the bandwidth efficiency of a shared mesh protection scheme is still largely unsolved. Ideally one would like analytical results that prove that (1) the optimal total bandwidth is never more than (say) twice the bandwidth that would be needed if no protection were required, and (2) there are polynomial time offline and online algorithms that are guaranteed to get within a certain percentage of the optimal total bandwidth. Unfortunately, these analytical results seem to be currently out of reach. Therefore it seems that the only thing to do is to obtain experimental results on specific instances. Although this is a reasonable approach, it raises the question, which instances should one choose?

There is no standard test suite for shared mesh protection studies. Most researchers therefore pick a few small networks more or less arbitrarily. This leaves one with the nagging worry that the instances chosen may not be truly representative of typical networks.

We do not have a fully satisfactory solution to this problem, but we suggest that some concepts from experimental design theory may be helpful. The idea is to identify certain parameters that are likely to be important and to determine the range of values that these parameters will take in practice. This defines a parameter space of possible instances. The parameter space will be too large to study exhaustively, so the goal is to sample it judiciously in order to obtain as much information as possible for the least possible effort.

For example, for shared mesh protection, three factors that seem to be important are (1) the average degree of the graph (this relates to how well-connected the graph is), (2) the girth (i.e., the length of the smallest cycle; since a working path and a protection path together form a cycle, a large girth means that short working paths will have long protection paths), and (3) the extent to which the demands are localized (i.e., whether each node wants to talk to everyone else or just to certain nodes). With these factors in mind, we have chosen six 12-node graphs, shown in Figure 4, to cover a range of possibilities in the parameter space. Tietze's graph, for example, is a small modification of the famous *Petersen graph*, which in turn is a *(3, 5)-cage*, i.e., the smallest regular graph of degree 3 and girth 5. The graph labeled "Murakami and Kim" is a slight modification of a network from [6]. The icosahedron has the maximum degree (five) attainable by a regular planar graph. Of course our selection of graphs is still somewhat arbitrary; it remains an open question how to apply experimental design methodology more rigorously in the selection process.

For each of the six graphs we have considered three different lists of demands: *uniform*, *nearest-neighbor*, and *unbalanced*. In uniform traffic, every pair of nodes appears exactly five times in the list of demands. In nearest-neighbor traffic, every pair of *adjacent* nodes appears exactly ten times in the list, and there are no other demands. In unbalanced traffic, three nodes are chosen to be "large" nodes and the remainder are deemed "small" nodes. The number of times a demand appears in the list depends on the "sizes" of the terminals: If $u$ and $v$ are respectively small/small, small/large, and large/large, then $\{u, v\}$ appears respectively 2, 8, and 14 times.

For each of the eighteen instances thus obtained, the demands were fed one at a time in a random order to our online algorithm and routed accordingly. For comparison, we also routed the demands using (1) a 1+1 path protection algorithm and (2) a simple shared path protection scheme that finds a node-disjoint pair of paths between every pair of terminals and routes all the demands on those paths, sharing when possible without worrying about branch points.

The results are shown in Table 1. The column labeled "Working" indicates the working bandwidth, which is the same for all three schemes—1+1, simple shared path, and PXT. The remaining three columns report the protection bandwidth for each of the three schemes; the column labeled "Path" refers to the simple shared path protection scheme. We see that our results confirm the conventional wisdom that shared mesh protection saves a lot of (total) bandwidth compared to dedicated protection. The percentage savings in our table varies from about 20% to 60%, which is a wider variation than has typically been reported in the literature; this may be due to our deliberate selection of a wide variety of networks. We also see that the bandwidth efficiency of the PXT algorithm is comparable to and often better than that of the conventional path protection algorithm, even though the PXT scheme is an online algorithm and the conventional path protection scheme is an offline algorithm. Therefore, the answer to our second question at the end of section 3 is that the no-branch-point constraint does not negate the bandwidth efficiency. This result is consonant with what Grover and Stamatelakis report for p-cycles.

Finally, we remark that we have run the PXT protection scheme on several much larger networks, including one with over two hundred nodes and over three hundred links, and with thousands of demands. The bandwidth efficiency was similar to that exhibited in Table 1, but unfortunately we cannot give further details, for intellectual property reasons.

## 6. CONCLUSION

The Achilles heel of shared mesh protection is its relatively slow restoration speed. This problem can be surmounted by forbidding branch points and thereby allowing protection paths to be pre-cross-connected. Grover and Stamatelakis took a first step in this direction with their p-cycle protection scheme, but we have gone further and have allowed a more general structure called a "pre-cross-connected trail" or PXT, whose flexibility allows it to be used easily in both link-based and path-based protection schemes and in both offline and online algorithms. Experimental results demonstrate that forbidding branch points does not destroy the main advantage of shared mesh protection, namely its high bandwidth efficiency.

## 7. APPENDIX : CONSTRAINED DIJKSTRA ALGORITHM

A key subroutine of our algorithm for computing allocation plans is a variant of Dijkstra's shortest-path algorithm that we call the *constrained Dijkstra algorithm*. Constrained Dijkstra is of some interest in its own right so we give a self-contained description of it in this section.

**INPUT.** A directed graph $G$, each of whose edges $e$ has a nonnegative weight (its *length*) and a (possibly empty) list of edges of $G$ (called the *rival edges* of $e$), and a distinguished node $v$ of $G$ (called the *source node*).

**OUTPUT.** For each node $u$ of $G$, the shortest *admissible* path from $v$ to $u$. A path $p$ is *admissible* if, for all edges $e$ in $p$, no rival edge of $e$ is in $p$.

**Definition.** A *partial path P* in $G$ is an ordered quadruple $(p, l, F, s)$, where $p$ is a directed path in $G$, $l$ is the length of the path (i.e., the sum of the lengths of its edges), $F$ is a set of edges of $G$ (called the *forbidden edges* of $P$), and $s$ is the *state* of the path (which takes one of two values: *penciled in* or *inked in*). We use the letters $p$, $l$ and $F$ to denote "coordinate functions," i.e., $F(P)$ is the set of forbidden edges of $P$, and so on.

**Definition.** A partial path $P_1$ is said to *dominate* a partial path $P_2$ if $l(P_1) \leq l(P_2)$ and $F(P_1) \subseteq F(P_2)$. Intuitively, this means that $P_1$ is at least as good as $P_2$. Note the use of $\leq$ rather than $<$ and $\subseteq$ rather than $\subset$. This is important.

**Preliminary remarks on the algorithm.** During the course of the algorithm, each node $u$ maintains a list of partial paths from $v$ to $u$. We say that a node is *black* if there exists an inked-in partial path in its list and we say that it is *white* otherwise. Initially only $v$ is black; as the algorithm runs, more and more white nodes become black. Once a node becomes black it stays black permanently.

If a node $u$ is black, it has at most one inked-in partial path; this represents the shortest admissible path from $v$ to $u$. If $u$ is white, its penciled-in partial paths represent paths that are potential shortest paths to $u$. If $u$ is black, its penciled-in partial paths represent initial segments of potential shortest paths to other nodes.

Like Dijkstra, constrained Dijkstra is a breadth-first search algorithm. At each step, one of the nodes $u$ is designated to be the *active node* and one of the partial paths of $u$ is designated to be the *active partial path*. Partial paths are extended one node at a time at the active node. Again like Dijkstra, constrained Dijkstra keeps the partial paths in a heap, so that it can quickly find the shortest partial path when it needs to.

**Initialization.** As a pre-processing step, we examine each edge $e$ of $G$ in turn; for each rival edge $f$ of $e$, we add $e$ to the list of rival edges of $f$ if $e$ is not already on that list. We are free to do this since it does not change the admissibility or length of any path in $G$, and it is convenient for our purposes.

The source node's list of partial paths is initialized to contain a single entry $P$: $p(P)$ is the path consisting solely of the source node $v$ itself, $l(P) = 0$, $F(P)$ is the empty set, and $s(P)$ has the value "inked in." Thus $v$ is black. We also designate $v$ to be the active node and its (unique) partial path to be the active partial path. At every other node the list of partial paths is empty, so they are all white. The partial path $P$ is put on a heap.

**Main loop.** We "probe forward" from the active node. That is, suppose that $u$ is the active node and that $P$ is the active partial path. We consider in turn each edge $e$ that emanates from $u$. If $e$ is forbidden, i.e., if $e \in F(P)$, then we ignore it and move on to the next edge. Otherwise, let $w$ be the node that $e$ points to. We let $P'$ be the partial path obtained from $P$ by appending $w$ to $p(P)$, adding the length of $e$ to $l(P)$, and adding the rival edges of $e$ to $F(P)$. If $P'$ is dominated by some partial path in $w$'s list, then we forget about it and move on to the next edge emanating from $u$. Otherwise, we add $P'$ to the list of partial paths at $w$, penciling it in. We also add it to the heap. We then delete any penciled-in partial paths in the list at $w$ that are dominated by $P'$. These partial paths are also deleted from the heap.

We repeat this process until all the edges emanating from *u* have been exhausted. We then remove *P* from the heap, but do not delete it from the list of partial paths at *u*.

Next, we extract the shortest partial path *Q* from the heap and designate it to be the new active partial path. We also designate the node *x* where we found *Q* to be the new active node. If *x* is white, we ink in *Q* (thereby making *x* black). Otherwise, we simply leave *Q* penciled in.

**Termination.** The algorithm terminates when we try to extract a partial path from the heap but find that it is empty, or when all nodes become black, whichever occurs first.

**Example.** Each edge in the directed graph in Figure 5 has three labels: the name of the edge, the length of the edge, and the set of rival edges. The distinguished node is $v_1$, which is the first black node.

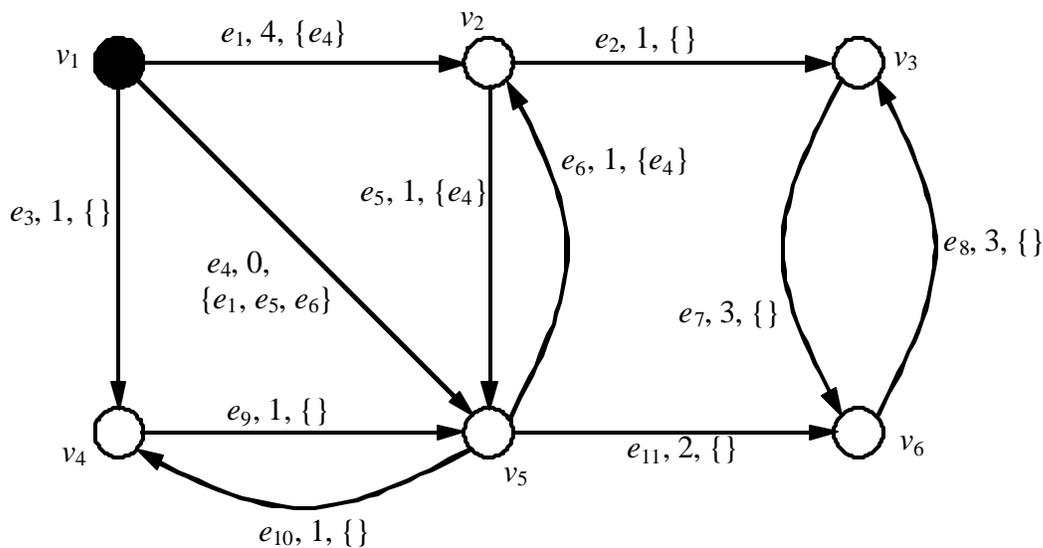

**Figure 5**

Observe that if we ignore the constraints given by the rival edges, then the shortest path from $v_1$ to $v_3$ is ($v_1$, $e_4$, $v_5$, $e_6$, $v_2$, $e_2$, $v_3$). However, this path is not admissible because it contains both $e_4$ and $e_6$, which are rivals of each other.

Initially $v_1$ is the active node. If we probe forward then we obtain three partial paths. The partial path at $v_5$ is the shortest so we ink it in, making $v_5$ black. These become the new active partial path and active node. (See Figure 6. To avoid clutter, we have suppressed edge labels. Shaded entries are penciled in and unshaded entries are inked in.)

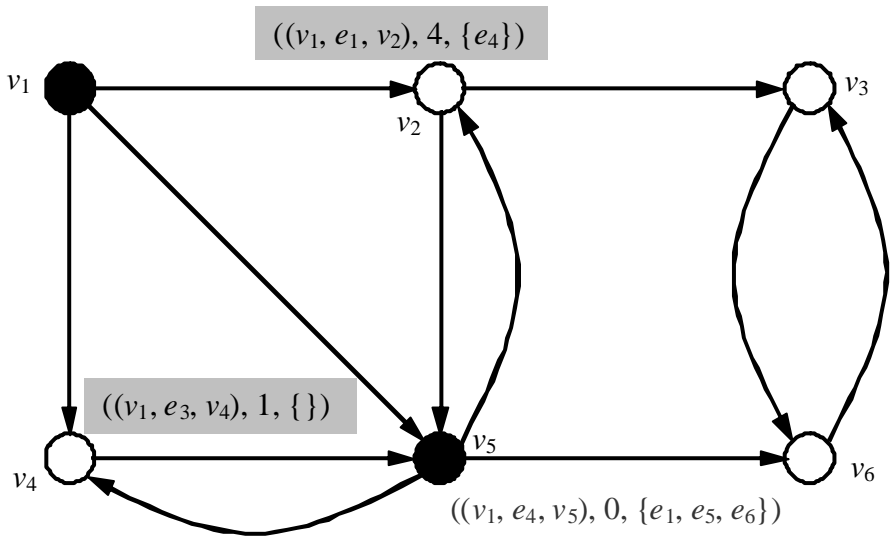

**Figure 6**

We now probe forward from $v_5$. At $v_4$, the new partial path is dominated by the existing partial path so it is not added. We cannot probe forward to $v_2$ because $e_6$ is forbidden. Probing forward to $v_6$ is all right and we add a new partial path there: $((v_1, e_4, v_5, e_{11}, v_6), 2, \{e_1, e_5, e_6\})$. This, however, does *not* become the new active partial path, because the penciled-in partial path at $v_4$ is shorter. We ink in the partial path at $v_4$, make $v_4$ black, and probe forward from $v_4$. The only new partial path created at this stage is at $v_5$: $((v_1, e_3, v_4, e_9, v_5), 1, \{\})$. Even though is $v_5$ is black, we retain this new partial path because it is not dominated by the existing partial path at $v_5$. (The existing partial path is shorter but has forbidden edges that are not forbidden in the new partial path.) In fact this becomes the new active partial path, although we do not ink it in because $v_5$ is already black.

Continuing in this way, we find that the remaining shortest admissible paths are $(v_1, e_3, v_4, e_9, v_5, e_6, v_2)$, $(v_1, e_3, v_4, e_9, v_5, e_6, v_2, e_2, v_3)$, and $(v_1, e_4, v_5, e_{11}, v_6)$. Notice that these paths do not arrange themselves into a tree; this is one difference from the usual Dijkstra algorithm.

**Final remarks.** We omit a detailed proof of the correctness of the constrained Dijkstra algorithm; the basic idea is that constrained Dijkstra is equivalent to ordinary Dijkstra on an auxiliary graph that can be built out of the partial paths.

Although we have described constrained Dijkstra for directed graphs, it can be applied to undirected graphs using the usual trick of replacing an undirected edge with two directed edges.

The running time of the constrained Dijkstra algorithm is exponential in the worst case. As an example of this, consider the "grid graph" $G_n$ whose nodes are the points in the plane whose coordinates are integers with absolute value at most $n$, and whose edges point from each vertex to its immediate southern neighbor and to its immediate western neighbor. Give each edge of $G_n$ one rival edge, namely its image under reflection in the line $x + y = 0$. It is not hard to show that if we take the node with coordinates $(n, n)$ as the source node, then by the time the algorithm first reaches the line $x + y = 0$ it will be keeping track of about $2^n$ partial paths.

Because of this potentially exponential consumption of resources, it is important that the actual implementation of the algorithm contain parameters that allow the algorithm to exit gracefully and report failure if it exceeds a certain amount of time or memory.

## 8. ACKNOWLEDGMENTS

Most of the work for this paper was done while the authors were at the Tellabs Research Center in Cambridge, Massachusetts. Tellabs Operations, Inc. has a patent pending on the methods in this paper. We thank James D. Mills, Philip J. Lin, Philippe Wilson, and Martin Lamothe for useful feedback.

| TABLE 1 | | | | |
|---|---|---|---|---|
|  | Working | 1+1 | Path | PXT |
| **UNIFORM** | | | | |
| **12-cycle + 3 edges** | 840 | 1440 | 905 | 894 |
| **3 x 4 grid** | 770 | 1070 | 495 | 587 |
| **Tietze's graph** | 645 | 1125 | 340 | 362 |
| **Murakami & Kim** | 600 | 820 | 560 | 533 |
| **Icosahedron** | 540 | 690 | 280 | 178 |
| **K6,6** | 480 | 840 | 365 | 139 |
|  | | | | |
| **NEIGHBOR** | | | | |
| **12-cycle + 3 edges** | 150 | 510 | 150 | 189 |
| **3 x 4 grid** | 170 | 510 | 170 | 236 |
| **Tietze's graph** | 180 | 690 | 170 | 206 |
| **Murakami & Kim** | 240 | 500 | 220 | 233 |
| **Icosahedron** | 300 | 600 | 290 | 205 |
| **K6,6** | 360 | 1080 | 200 | 188 |
|  | | | | |
| **UNBALANCED** | | | | |
| **12-cycle + 3 edges** | 768 | 1368 | 824 | 794 |
| **3 x 4 grid** | 704 | 1004 | 594 | 476 |
| **Tietze's graph** | 636 | 1152 | 436 | 395 |
| **Murakami & Kim** | 516 | 742 | 450 | 399 |
| **Icosahedron** | 540 | 690 | 356 | 210 |
| **K6,6** | 480 | 840 | 378 | 154 |

3 x 4

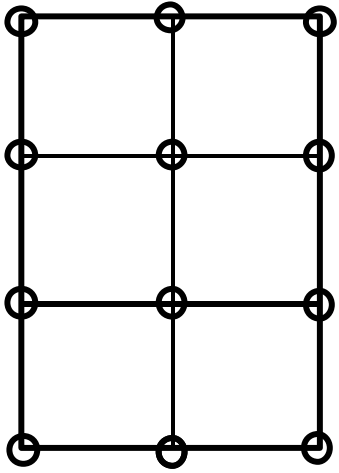

Icosahedron

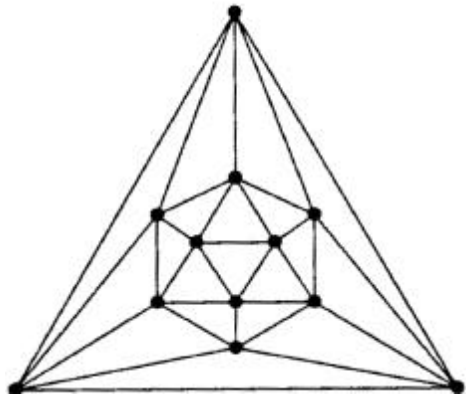

12-cycle + 3 edges

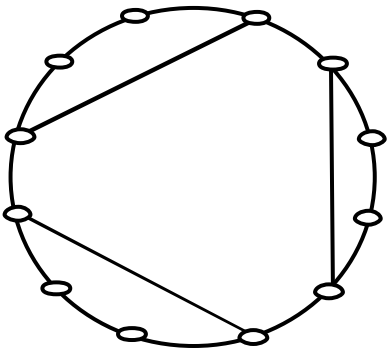

$K_{6,6}$

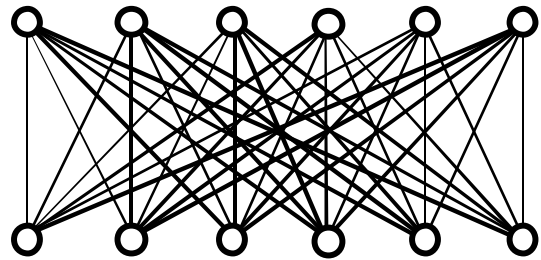

Tietze's Graph

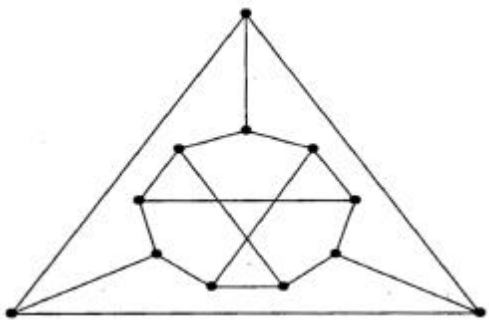

Murakami & Kim

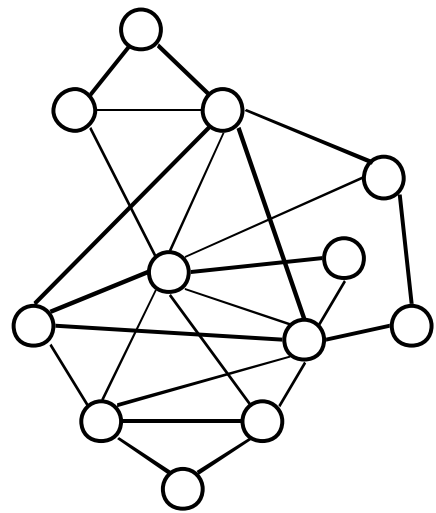

**Figure 4**